\documentclass[aps,prl,twocolumn,groupedaddress,showpacs,showkeys]{revtex4}
\usepackage{graphicx,psfrag,here,epsfig}

\begin{document}
\def\eq{\begin{eqnarray}}
\def\en{\end{eqnarray}}
\hyphenation{thres-hold}

\preprint{HISKP-TH-05/22, FZJ-IKP-TH-2005-36}

\title{Isospin-breaking corrections in the pion-deuteron scattering length}

\author{Ulf-G. Mei{\ss}ner}
\email[]{meissner@itkp.uni-bonn.de}
\affiliation{Universit\"{a}t Bonn, Helmholtz-Institut f\"{u}r
Strahlen- und Kernphysik (Theorie), Nu\ss allee 14-16, D-53115 Bonn, Germany,
and Forschungszentrum J\"{u}lich, Institut f\" {u}r Kenphysik (Theorie),
D-52425 J\"{u}lich, Germany}

\author{Udit Raha}
\email[]{udit@itkp.uni-bonn.de}
\affiliation{Universit\"{a}t Bonn, Helmholtz-Institut f\"{u}r
Strahlen- und Kernphysik (Theorie), Nu\ss allee 14-16, D-53115 Bonn, Germany}

\author{Akaki Rusetsky}
\email[]{rusetsky@itkp.uni-bonn.de}
\affiliation{Universit\"{a}t Bonn, Helmholtz-Institut f\"{u}r
Strahlen- und Kernphysik (Theorie), Nu\ss allee 14-16, D-53115 Bonn, Germany,
and High Energy 
Physics Institute, Tbilisi State University, University St.~9, 380086
 Tbilisi, Georgia}

\date{\today}

\begin{abstract}
It is shown that isospin-breaking corrections to the pion-deuteron
scattering length can be very large, because of the vanishing of the 
isospin-symmetric contribution to this scattering length at leading order
in chiral perturbation theory. We further demonstrate that these corrections 
can explain the bulk of the discrepancy between the recent experimental
data on pionic hydrogen and pionic deuterium. We also give the first determination
of the electromagnetic low-energy constant~$f_1$.
\end{abstract}


\maketitle

Already in 1977, Weinberg pointed out~\cite{Weinberg:1977hb}
that the isospin-breaking corrections to certain pion-nucleon scattering
amplitudes could become large because the iso\-spin-sym\-met\-ric
contributions to these amplitudes are chirally suppressed. Unfortunately,
Weinberg's statement refers to the scattering processes with neutral pions
that makes its difficult to verify with present experimental techniques. 
It turns out, however,
that there exists a fascinating possibility to directly 
observe a large isospin-breaking
correction in the negatively
charged pion elastic scattering on the deuteron, where the 
leading-order isospin-symmetric amplitude in chiral perturbation theory
(ChPT)
is proportional to the isospin-even pion-nucleon scattering length $a^+$
and is thus very small. Quite surprisingly, such (a rather obvious) phenomenon
has not been explored so far.
Studies of  isospin breaking in the $\pi d$ system 
(see, e.g. \cite{Deloff:2001zp,Doring:2004kt})
include effects coming from the
Coulomb field and/or the particle mass differences in the loops. 
Numerically these effects indeed turn out to be moderate. 
However, as it is well known (see 
e.g.~\cite{Meissner:1997ii,Fettes:1998wf,Fettes:2001cr,Lyubovitskij:2000kk,Gasser:2002am}), 
isospin breaking in ChPT at leading order emerges through the direct 
quark-photon coupling encoded in the
electromagnetic low-energy constants (LECs) of the effective chiral Lagrangian,
as well as due to the explicit quark mass dependence of the pion-nucleon
amplitudes.  To the best of our knowledge, neither of these mechanisms have
been taken into account in the existing investigations on pionic deuterium.
  
On the other hand, the presence of large isospin-breaking corrections may
have serious implications for the combined analysis of the experimental data
on pionic hydrogen and pionic deuterium, which is aimed at the precise
determination the $S$-wave pion-nucleon scattering 
lengths \cite{Schroder:2001rc,Hauser:1998yd,Gotta:2005cg} 
(see also \cite{Ericson:2000md}).
As it is well known,
in the experiment one measures the ground-state energy shift 
($\epsilon_{1s}$) and the width ($\Gamma_{1s}$) of  pionic hydrogen, as well as
the ground-state energy shift of  pionic deuterium ($\epsilon_{1s}^d$).
Applying a theoretically calculated set of the isospin-breaking corrections,
the hydrogen observables are related to the isospin-even and -odd $S$-wave
$\pi N$ scattering lengths $a^+$ and $a^-$ 
\eq
\epsilon_{1s}&=&-2\alpha^3\mu_r^2(a^++a^-)(1+\delta_\epsilon)\, ,
\nonumber\\[2mm]
\Gamma_{1s}&=&8\alpha^3\mu_r^2p_0\biggl(1+\frac{1}{P}\biggr)
\left(a^-(1+\delta_\Gamma)\right)^2\, ,
\en
whereas the real part of the pion-deuteron scattering length $a_{\pi d}$
is expressed through the measured shift in  pionic deuterium 
\eq
\epsilon_{1s}^d&=&-2\alpha^3\mu_d^2\,{\rm Re}\, a_{\pi d}\, .
\en
In the above formulae, $\mu_r$ and $\mu_d$ denote the reduced mass of the 
$\pi^-p$ and $\pi^-d$ systems, respectively, $P=1.546\pm 0.009$ is the Panofsky
ratio, $\alpha$ the fine-structure constant, $p_0$ is the center-of-mass 
momentum of the $\pi^0n$ pair, which emerges in 
the decay of the pionic hydrogen
and $\delta_\epsilon$, $\delta_\Gamma$ stand for the 
isospin-breaking corrections. 

\begin{figure*}
 \includegraphics[width=11.4cm]{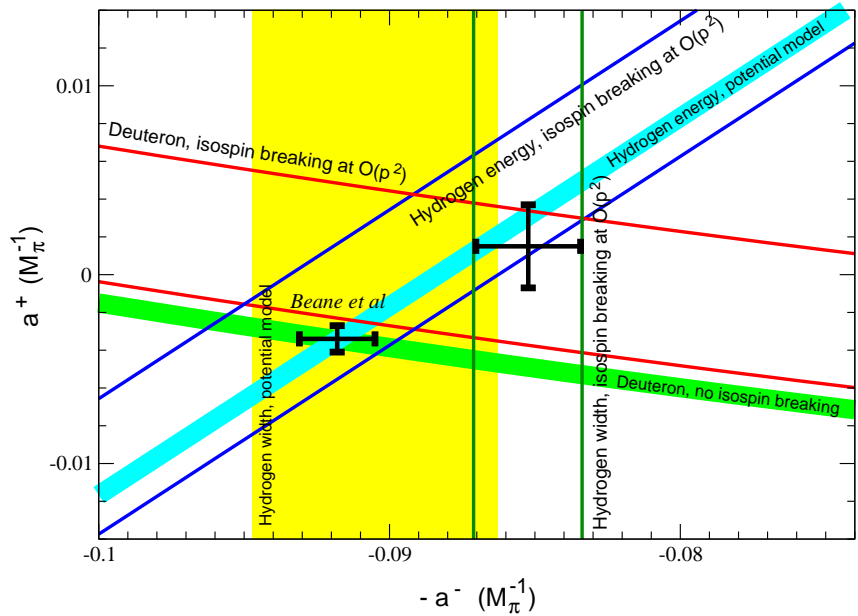}%
 \caption{Determination of the $\pi N$ $S$-wave scattering lengths
$a^+$ and $a^-$ from the combined analysis of the experimental 
data on the pionic hydrogen
energy shift and width, as well as
the pionic deuterium energy shift (details in the text).
The cross denoted as {\it Beane et al} is taken from 
Ref.~\cite{Beane:2002wk}. The second cross corresponds to
the scattering lengths given in Eq.~(\ref{eq:scattl}).
\label{fig:plot}}
 \end{figure*}

The present status of the data analysis, which is performed by the
 Pionic Hydrogen Collaboration at PSI,  is the following.
Until very recently, one  used 
a set of  corrections calculated using a potential
model approach \cite{Sigg:1996qe}: $\delta_\epsilon=(-2.1\pm 0.5)\cdot 10^{-2}$ and
$\delta_\Gamma=(-1.3\pm 0.5)\cdot 10^{-2}$, with
$\epsilon_{1s}=-7.108\pm 0.013\pm 0.034~{\rm eV}$ and 
$\Gamma_{1s}=0.868\pm 0.040\pm 0.038~{\rm eV}$, 
whereas the isospin-breaking corrections in the deuteron have been 
neglected~\cite{Schroder:2001rc}. 
This led to a coherent picture which is
shown in Fig.~\ref{fig:plot}. In this plot, the constraints on the
$a^+$ and $a^-$, which emerge
from these three different measurements, are shown as three shaded strips
denoted as ``Hydrogen energy, potential model,''
``Hydrogen width, potential model'' and ``Deuteron, no isospin breaking.'' 
The first strip corresponds to the value of
$a^+ + a^-$ determined from the energy shift measurement in pionic hydrogen, 
the second strip defines $a^-$ from the 
hydrogen width and the third strip
is obtained from the experimental  value  
${\rm Re}\,a_{\pi d}^{\rm exp}=-(0.0261\pm 0.0005)M_\pi^{-1}$
\cite{Hauser:1998yd} by applying the formula
\eq\label{eq:bernard}
\hspace*{-.1cm}&& {\rm Re}\,\bar a_{\pi d}=2\,\frac{1+\mu}{1+\mu/2}\,a^+
\nonumber\\
\hspace*{-.1cm}&+&2\,\frac{(1+\mu)^2}{1+\mu/2}\,\left((a^+)^2-2(a^-)^2\right)\,
\frac{1}{2\pi^2}\,
\left\langle\frac{1}{{\bf q}^2}\right\rangle_{\rm wf}
\nonumber\\
\hspace*{-.1cm}&+&2\,\frac{(1+\mu)^3}{1+\mu/2}\,\left((a^+)^3-2(a^-)^2(a^+-a^-)\right)
\,\frac{1}{4\pi}\,
\left\langle\frac{1}{|{\bf q}|}\right\rangle_{\rm wf}
\nonumber\\
\hspace*{-.1cm}&+&a_{\rm boost}+\cdots\, ,
\en
which was derived in Ref.~\cite{Beane:2002wk} within  ChPT
under the assumption of  exact isospin symmetry (the bar over
$a_{\pi d}$ refers to the quantities in the isospin limit).
In the above equation, $a_{\rm boost}=(0.00369\cdots 0.00511)M_\pi^{-1}$
and $\left\langle 1 / {\bf q}^2 \right\rangle_{\rm wf} = (12.3\pm 0.3)M_\pi\,$
and $\left\langle 1 /|{\bf q}|\right\rangle_{\rm wf}
= (7.2\pm 1.0) M_\pi^2\,$,
where  NLO wave functions with the cutoff mass in the interval
$\Lambda=(500\cdots 600)~{\rm MeV}$  \cite{Epelbaum:1999dj}
have been used in order to evaluate
the above wave-function averages. As one immediately sees from the plot, all
three strips intersect in a small domain of the $(a^+,a^-)$-plane shown by a 
cross. The resulting values of the
scattering lengths are $a^+=(-0.0034\pm 0.0007)M_\pi^{-1}$ and 
$a^-=(0.0918\pm 0.0013) M_\pi^{-1}$~\cite{Beane:2002wk}.

Recent developments have completely changed the picture. On the experimental
side, new measurements have resulted not only in a considerably improved 
accuracy for the hydrogen width, but in a shift of its central value as well.
The latest results are \cite{Gotta:2005cg}:
$\epsilon_{1s}=-7.120\pm 0.008\pm 0.009~{\rm eV}$ and
$\Gamma_{1s}=0.800\pm 0.030~{\rm eV}$. On the theoretical side, there now
exist calculations of the isospin-breaking corrections in ChPT
at $O(p^2)$~\cite{Lyubovitskij:2000kk,Gasser:2002am,Zemp}. 
Applying these 
corrections to the latest experimental data, we get two shifted bands which are
shown in Fig.~\ref{fig:plot}. Note also that the increase of the uncertainty
in the energy shift in \cite{Lyubovitskij:2000kk,Gasser:2002am} was 
largely due to the unknown $O(p^2)$ LEC 
$f_1$, which was first considered
in~\cite{Meissner:1997ii} and  
merely omitted in  the potential model calculations. 
In the case of the energy shift, the correction at
$O(p^3)$ is available as well~\cite{Gasser:2002am}. For consistency reasons
however, we use $O(p^2)$ results everywhere. As evident from the plot,
the three bands do not have a common intersection domain any more.
It could of course be argued that the discrepancy is due to the incomplete
treatment of the deuteron structure. Recent investigations
(see, e.g. 
\cite{Beane:2002aw,Beane:2002wk,Borasoy:2003gf,Meissner:2005bz,Nogga:2005fv}),
however, converge to the conclusion that the uncertainty in the three-body
calculations can be made rather small -- at least, it can not
be  solely responsible for the large discrepancy which one observes in 
Fig.~\ref{fig:plot}. With this conclusion, 
it becomes evident that we encounter a serious problem
in the interpretation of the experimental data.

The only loophole left in the theoretical treatment of the $\pi d$ 
scattering length lies in the assumption of exact isospin invariance. 
In the real world  isospin symmetry 
is broken due to two distinct sources: the electromagnetic
interactions and the $u,d$ quark mass difference. It is convenient to
to treat these two effects on the same footing, introducing a formal 
parameter $\delta$ so that $\alpha\sim (m_d-m_u)\sim\delta$. In ChPT
the parameter $\delta$ is counted
as $O(p^2)$. Further, in the isospin-symmetric world, by convention, the
masses of the pions and the nucleons coincide with the charged pion mass
$M_{\pi^\pm}\doteq M_\pi$ and the proton mass $m_p$, 
respectively (see e.g. Ref.~\cite{Gasser:2003hk} for more details).
The $\pi d$ scattering length is given by
\eq
{\rm Re}\,a_{\pi d}&=&{\rm Re}\,\bar a_{\pi d}+\Delta a_{\pi d}\, ,
\nonumber\\
\Delta a_{\pi d}&=&A_1\alpha+A_2(m_d-m_u)+O(\delta^2)\, ,
\en
and $A_1,A_2$ can be further expanded in quark masses in ChPT.
The expansion of the isospin-breaking contribution starts at order $p^2$
\eq
\Delta a_{\pi d}&=&\Delta a_{\pi d}^{\rm LO}+O(p^3)\, ,
\en
i.e. at the same order as the leading isospin-conserving term in 
Eq.~(\ref{eq:bernard}). For this reason, it is  no wonder that the 
isospin-breaking corrections turn out to be large.


Let us now consider different pion-nucleon scattering amplitudes
in the physical particle basis:
$\pi^-p\to\pi^-p$ (labeled hereafter as ``$p$''),
$\pi^-n\to\pi^-n$ (``$n$'') and $\pi^-p\to\pi^0n$ (``$x$''). 
In the absence of  virtual photons, the threshold amplitudes are defined
according to
\eq
\langle \pi'N'|T|\pi N\rangle\bigr|_{\rm thr}=
\bar u(N'){\cal T}_{p,n,x}u(N)\, ,
\en
where we use the normalization $\bar u(N)u(N)=2m_N$ and $N$ 
denotes the proton or the neutron. In the presence of  virtual photons,
the definition of the threshold amplitude is modified: prior to approaching 
the threshold, one has to subtract one-photon exchange contribution and to
remove the (infrared-divergent) Coulomb 
phase~\cite{Lyubovitskij:2000kk,Gasser:2002am}.
Below we do not consider these complications, because we work 
at $O(p^2)$ in ChPT and the virtual photons are still 
not present at this order. Further, using the Condon-Shortley phase 
convention, we may write
\eq
{\cal T}_{p}&=&4\pi(1+\mu)(a^+ + a^-)+\delta {\cal T}_p\, ,
\nonumber\\
{\cal T}_n&=&4\pi(1+\mu)(a^+ - a^-)+\delta {\cal T}_n\, ,
\nonumber\\
{\cal T}_x&=&4\pi(1+\mu)(\sqrt{2}a^-)+\delta {\cal T}_x\, ,
\en
where $\mu=M_\pi/m_p$ and $\delta {\cal T}_{p,n,x}=O(\delta)$.
At $O(p^2)$ in ChPT we 
get~\cite{Fettes:1998wf,Fettes:2001cr,Lyubovitskij:2000kk,Gasser:2002am,Zemp}
\eq
\delta{\cal T}_p&=&\frac{4(M_\pi^2-M_{\pi^0}^2)}{F_\pi^2}\,c_1
-\frac{e^2}{2}\,(4f_1+f_2)+O(p^3)\, ,
\nonumber\\
\delta{\cal T}_n&=&\frac{4(M_\pi^2-M_{\pi^0}^2)}{F_\pi^2}\,c_1
-\frac{e^2}{2}\,(4f_1-f_2)+O(p^3)\, ,
\nonumber\\
\delta{\cal T}_x&=&
-\sqrt{2}\biggl(\frac{g_A^2(M_\pi^2-M_{\pi^0}^2)}{4m_pF_\pi^2}
+\frac{e^2f_2}{2}\biggr)+O(p^3)\, ,
\en
where $F_\pi=92.4~{\rm MeV}$ is the pion decay constant,
$g_A=1.27$ denotes the axial-vector charge of a nucleon and
$c_1$ and $f_1,f_2$ stand for the $O(p^2)$ 
strong and electromagnetic LECs, respectively.
In the numerical calculations we take
$c_1=-0.9^{+0.5}_{-0.2}~{\rm GeV}^{-1}$~\cite{Meissner:2005ba},
$f_2=-(0.97\pm 0.38)~{\rm GeV}^{-1}$ \cite{Gasser:2002am}. 
Note that the errors on the LEC $c_1$ are most conservative.
The largest uncertainty in the
results is introduced by the constant $f_1$, whose value at present
is unknown and for which  the dimensional estimate
$|f_1|\leq 1.4~{\rm GeV}^{-1}$ has been used. Note also, that the hydrogen
energy band, which is shown in Fig.~\ref{fig:plot}
corresponds to the new value of $c_1$ given above.

At the leading order, the isospin-breaking correction to the $\pi d$
scattering length is given by
\eq\label{eq:LO}
\Delta a_{\pi d}^{\rm LO}&=&(4\pi(1+\mu/2))^{-1}(\delta {\cal T}_p
+\delta {\cal T}_n)\, .
\en
One sees that the leading-order isospin-breaking correction
is independent
on the deuteron structure, which enters in the subsequent terms through
the wave-function averages. 
Substituting numerical values for the various low-ener\-gy constants
one obtains that the correction is extremely large
\begin{equation}
\Delta a_{\pi d}^{\rm LO} = -(0.0110^{+0.0081}_{-0.0058}) \, M_\pi^{-1}~,
\end{equation}
that is
$\Delta a_{\pi d}^{\rm LO}/ {\rm Re}\,a_{\pi d}^{\rm exp}=0.42$ 
(central values). Moreover, one can immediately see that
the correction moves the de\-u\-te\-ron band in 
Fig.~\ref{fig:plot} in the right direction: the isospin-breaking 
corrections amount for the bulk of the discrepancy between the experimental
data on  pionic hydrogen and deuterium. 
Including the corrections $\Delta a_{\pi d}^{\rm LO}$, all bands now 
have a common 
intersection area in the $a^+,a^-$-plane, see Fig.~\ref{fig:plot}.
The resulting values for the $\pi N$ scattering lengths are:
\begin{eqnarray}\label{eq:scattl}
a^+ &=& (0.0015\pm 0.0022) \, M_\pi^{-1}\, , \nonumber\\ 
a^- &=& (0.0852\pm 0.0018) \, M_\pi^{-1}\, .  
\end{eqnarray}
Further, using the hydrogen energy shift to estimate the LEC $f_1$, we obtain
\begin{equation}
f_1= -2.1^{+3.2}_{-2.2}~{\rm GeV}^{-1}\, .
\end{equation}
Note that the error displayed here does not include the uncertainty coming
from the higher orders in ChPT and should thus be considered preliminary. 
We also wish to point out that 
the central value of $f_1$ (large and negative) agrees with a recent
model-based estimate~\cite{Lyubovitskij:2001zn}.

As we see, the presence of the $O(p^2)$ LECs in the expressions for the
isospin-breaking corrections leads to a sizeable increase of the uncertainty
in the output. In order to gain  precision, 
in the fit one might also use those particular linear
combination(s) of the experimental observables that 
do not contain $f_1$ and $c_1$. 
To carry out such a combined analysis with the required precision
one would first have to evaluate the  isospin-breaking corrections 
with a better accuracy.

Up to now, we have restricted ourselves to the leading-order isospin-breaking
correction in ChPT. Below, we briefly consider higher orders, where
a full-fledged investigation has not been done yet.
Calculations at $O(p^3)$ exist only for the hydrogen
energy shift and yield 
$\delta_\epsilon=(-7.2\pm 2.9)\cdot 10^{-2}$~\cite{Gasser:2002am} (using
$c_1=(-0.93\pm 0.07)~{\rm GeV}^{-1}$).
The corrections to $O(p^2)$ result are sizable
(the energy band in Fig.~\ref{fig:plot} will be shifted further upwards), 
but the uncertainty,
which is almost completely determined by the $O(p^2)$ LECs, 
remains practically the same.
On the other hand, consistent studies at $O(p^3)$ 
imply the treatment of the scattering process in the
three-body system in the effective field theory with virtual photons. 
To the best of our knowledge, such investigations have not been yet carried 
out. At the moment, it would be plausible to put forward the conjecture that
all isospin-breaking effects in 
${\rm Re}\,a_{\pi d}$ at $O(p^3)$ still emerge from the first term in the
multiple-scattering series, which contains only ${\cal T}_p$ and ${\cal T}_n$,
whereas the corrections that depend on the structure of the deuteron,
start at $O(p^4)$. From this we expect that in order to extract
more precise experimental values of the scattering lengths, it might suffice
to obtain a full set of isospin-breaking corrections to the $\pi N$ amplitudes
$\delta {\cal T}_{p,n,x}$ at $O(p^3)$ in ChPT. Of course, the discussion 
given here can not be a substitute for a rigorous proof in the framework
of EFT, which in the light of the above discussion, is urgently needed.

As mentioned above, isospin-breaking corrections at $O(p^4)$ depend on the
details of the $NN$ interactions and the deuteron structure. In practice,
it might prove rather difficult to evaluate these corrections with sufficient
accuracy. On the other hand, in order to get a feeling of
how large the  $O(p^4)$-contributions could be, let us consider a typical
 correction which emerges
from the double-scattering term in the multiple-scattering series, 
Eq.~(\ref{eq:bernard})
\eq\label{eq:double}
\hspace*{-.2cm}&&\Delta a_{\pi d}^{\rm double~scat.}
=\frac{1+\mu}{4\pi^3(1+\mu/2)}\,
\left\langle\frac{1}{{\bf q}^2}\right\rangle_{\rm wf}
\nonumber\\
\hspace*{-.2cm}&\times&\biggl\{ (a^+-a^-)\delta{\cal T}_p+(a^++a^-)\delta{\cal T}_n
-\sqrt{2}a^-\delta{\cal T}_x\biggr\}
\\
\hspace*{-.2cm}&=&(-0.023+0.028-0.002)\,{\rm Re}\,a_{\pi d}^{\rm exp}
=0.003\, {\rm Re}\,a_{\pi d}^{\rm exp}\, .
\nonumber
\en
As an input, we have used the scattering lengths from Eq.~(\ref{eq:scattl}).
The individual terms in this equation amount to
a few percent of the isospin-symmetric contribution
(with additional cancellations in the sum) that 
may serve as a rough order-of-magnitude estimate for the isospin-breaking
corrections at $O(p^4)$.

Last but not least, we note that in our opinion, the above discussion
 clearly justifies the need 
for an improved measurement of the energy shift in  pionic deuterium.
As we have seen, the uncertainties, which emerge in the treatment of the 
deuteron structure within 
effective field theory are much smaller than the uncertainties due to the
presence of the $O(p^2)$ LECs (although, there is
still some room for the improved higher-order calculations in the 
isospin-symmetric sector). On the other hand, it is evident that a major
effort is needed on the theoretical side: in order to extract the scattering
lengths to a good precision, the isospin-breaking corrections 
should be evaluated at least at $O(p^3)$ in ChPT.

\begin{acknowledgments}
The authors would like to thank J. Gasser, C. Hanhart and A. Nogga
for very interesting discussions.
Partial financial support under the EU Integrated Infrastructure
Initiative Hadron Physics Project (contract number RII3-CT-2004-506078)
and DFG (SFB/TR 16, ``Subnuclear Structure of Matter'') is gratefully
acknowledged.
\end{acknowledgments}

\bibliography{isospin_breaking}

\begin{thebibliography}{23}
\expandafter\ifx\csname natexlab\endcsname\relax\def\natexlab#1{#1}\fi
\expandafter\ifx\csname bibnamefont\endcsname\relax
  \def\bibnamefont#1{#1}\fi
\expandafter\ifx\csname bibfnamefont\endcsname\relax
  \def\bibfnamefont#1{#1}\fi
\expandafter\ifx\csname citenamefont\endcsname\relax
  \def\citenamefont#1{#1}\fi
\expandafter\ifx\csname url\endcsname\relax
  \def\url#1{\texttt{#1}}\fi
\expandafter\ifx\csname urlprefix\endcsname\relax\def\urlprefix{URL }\fi
\providecommand{\bibinfo}[2]{#2}
\providecommand{\eprint}[2][]{\url{#2}}

\bibitem[{\citenamefont{Weinberg}(1977)}]{Weinberg:1977hb}
\bibinfo{author}{\bibfnamefont{S.}~\bibnamefont{Weinberg}},
  \bibinfo{journal}{Trans. New York Acad. Sci.} \textbf{\bibinfo{volume}{38}},
  \bibinfo{pages}{185} (\bibinfo{year}{1977}).

\bibitem[{\citenamefont{Deloff}(2001)}]{Deloff:2001zp}
\bibinfo{author}{\bibfnamefont{A.}~\bibnamefont{Deloff}},
  \bibinfo{journal}{Phys. Rev.} \textbf{\bibinfo{volume}{C64}},
  \bibinfo{pages}{065205} (\bibinfo{year}{2001}), \eprint{nucl-th/0104067}.

\bibitem[{\citenamefont{D{\"o}ring et~al.}(2004)\citenamefont{D{\"o}ring, Oset,
  and Vicente~Vacas}}]{Doring:2004kt}
\bibinfo{author}{\bibfnamefont{M.}~\bibnamefont{D{\"o}ring}},
  \bibinfo{author}{\bibfnamefont{E.}~\bibnamefont{Oset}}, \bibnamefont{and}
  \bibinfo{author}{\bibfnamefont{M.~J.} \bibnamefont{Vicente~Vacas}},
  \bibinfo{journal}{Phys. Rev.} \textbf{\bibinfo{volume}{C70}},
  \bibinfo{pages}{045203} (\bibinfo{year}{2004}), \eprint{nucl-th/0402086}.

\bibitem[{\citenamefont{Mei{\ss}ner and Steininger}(1998)}]{Meissner:1997ii}
\bibinfo{author}{\bibfnamefont{U.-G.} \bibnamefont{Mei{\ss}ner}}
  \bibnamefont{and}
  \bibinfo{author}{\bibfnamefont{S.}~\bibnamefont{Steininger}},
  \bibinfo{journal}{Phys. Lett.} \textbf{\bibinfo{volume}{B419}},
  \bibinfo{pages}{403} (\bibinfo{year}{1998}), \eprint{hep-ph/9709453}.

\bibitem[{\citenamefont{Fettes et~al.}(1999)\citenamefont{Fettes, Mei{\ss}ner,
  and Steininger}}]{Fettes:1998wf}
\bibinfo{author}{\bibfnamefont{N.}~\bibnamefont{Fettes}},
  \bibinfo{author}{\bibfnamefont{U.-G.} \bibnamefont{Mei{\ss}ner}},
  \bibnamefont{and}
  \bibinfo{author}{\bibfnamefont{S.}~\bibnamefont{Steininger}},
  \bibinfo{journal}{Phys. Lett.} \textbf{\bibinfo{volume}{B451}},
  \bibinfo{pages}{233} (\bibinfo{year}{1999}), \eprint{hep-ph/9811366}.

\bibitem[{\citenamefont{Fettes and Mei{\ss}ner}(2001)}]{Fettes:2001cr}
\bibinfo{author}{\bibfnamefont{N.}~\bibnamefont{Fettes}} \bibnamefont{and}
  \bibinfo{author}{\bibfnamefont{U.-G.} \bibnamefont{Mei{\ss}ner}},
  \bibinfo{journal}{Nucl. Phys.} \textbf{\bibinfo{volume}{A693}},
  \bibinfo{pages}{693} (\bibinfo{year}{2001}), \eprint{hep-ph/0101030}.

\bibitem[{\citenamefont{Lyubovitskij and Rusetsky}(2000)}]{Lyubovitskij:2000kk}
\bibinfo{author}{\bibfnamefont{V.~E.} \bibnamefont{Lyubovitskij}}
  \bibnamefont{and} \bibinfo{author}{\bibfnamefont{A.}~\bibnamefont{Rusetsky}},
  \bibinfo{journal}{Phys. Lett.} \textbf{\bibinfo{volume}{B494}},
  \bibinfo{pages}{9} (\bibinfo{year}{2000}), \eprint{hep-ph/0009206}.

\bibitem[{\citenamefont{Gasser et~al.}(2002)\citenamefont{Gasser, Ivanov,
  Lipartia, Moj{\v z}i{\v s}, and Rusetsky}}]{Gasser:2002am}
\bibinfo{author}{\bibfnamefont{J.}~\bibnamefont{Gasser}},
  \bibinfo{author}{\bibfnamefont{M.~A.} \bibnamefont{Ivanov}},
  \bibinfo{author}{\bibfnamefont{E.}~\bibnamefont{Lipartia}},
  \bibinfo{author}{\bibfnamefont{M.}~\bibnamefont{Moj{\v z}i{\v s}}},
  \bibnamefont{and} \bibinfo{author}{\bibfnamefont{A.}~\bibnamefont{Rusetsky}},
  \bibinfo{journal}{Eur. Phys. J.} \textbf{\bibinfo{volume}{C26}},
  \bibinfo{pages}{13} (\bibinfo{year}{2002}), \eprint{hep-ph/0206068}.

\bibitem[{\citenamefont{Schr{\"o}der et~al.}(2001)}]{Schroder:2001rc}
\bibinfo{author}{\bibfnamefont{H.~C.} \bibnamefont{Schr{\"o}der}}
  \bibnamefont{et~al.}, \bibinfo{journal}{Eur. Phys. J.}
  \textbf{\bibinfo{volume}{C21}}, \bibinfo{pages}{473} (\bibinfo{year}{2001}).

\bibitem[{\citenamefont{Hauser et~al.}(1998)}]{Hauser:1998yd}
\bibinfo{author}{\bibfnamefont{P.}~\bibnamefont{Hauser}} \bibnamefont{et~al.},
  \bibinfo{journal}{Phys. Rev.} \textbf{\bibinfo{volume}{C58}},
  \bibinfo{pages}{1869} (\bibinfo{year}{1998}).

\bibitem[{\citenamefont{Gotta}(2005)}]{Gotta:2005cg}
\bibinfo{author}{\bibfnamefont{D.}~\bibnamefont{Gotta}}
  (\bibinfo{collaboration}{Pionic Hydrogen Collaboration}),
  \bibinfo{journal}{Int. J. Mod. Phys.} \textbf{\bibinfo{volume}{A20}},
  \bibinfo{pages}{349} (\bibinfo{year}{2005}).

\bibitem[{\citenamefont{Ericson et~al.}(2002)\citenamefont{Ericson, Loiseau,
  and Thomas}}]{Ericson:2000md}
\bibinfo{author}{\bibfnamefont{T.~E.~O.} \bibnamefont{Ericson}},
  \bibinfo{author}{\bibfnamefont{B.}~\bibnamefont{Loiseau}}, \bibnamefont{and}
  \bibinfo{author}{\bibfnamefont{A.~W.} \bibnamefont{Thomas}},
  \bibinfo{journal}{Phys. Rev.} \textbf{\bibinfo{volume}{C66}},
  \bibinfo{pages}{014005} (\bibinfo{year}{2002}), \eprint{hep-ph/0009312}.

\bibitem[{\citenamefont{Sigg et~al.}(1996)\citenamefont{Sigg, Badertscher,
  Goudsmit, Leisi, and Oades}}]{Sigg:1996qe}
\bibinfo{author}{\bibfnamefont{D.}~\bibnamefont{Sigg}},
  \bibinfo{author}{\bibfnamefont{A.}~\bibnamefont{Badertscher}},
  \bibinfo{author}{\bibfnamefont{P.~F.~A.} \bibnamefont{Goudsmit}},
  \bibinfo{author}{\bibfnamefont{H.~J.} \bibnamefont{Leisi}}, \bibnamefont{and}
  \bibinfo{author}{\bibfnamefont{G.~C.} \bibnamefont{Oades}},
  \bibinfo{journal}{Nucl. Phys.} \textbf{\bibinfo{volume}{A609}},
  \bibinfo{pages}{310} (\bibinfo{year}{1996}).

\bibitem[{\citenamefont{Beane et~al.}(2003)\citenamefont{Beane, Bernard,
  Epelbaum, Mei{\ss}ner, and Phillips}}]{Beane:2002wk}
\bibinfo{author}{\bibfnamefont{S.~R.} \bibnamefont{Beane}},
  \bibinfo{author}{\bibfnamefont{V.}~\bibnamefont{Bernard}},
  \bibinfo{author}{\bibfnamefont{E.}~\bibnamefont{Epelbaum}},
  \bibinfo{author}{\bibfnamefont{U.-G.} \bibnamefont{Mei{\ss}ner}},
  \bibnamefont{and} \bibinfo{author}{\bibfnamefont{D.~R.}
  \bibnamefont{Phillips}}, \bibinfo{journal}{Nucl. Phys.}
  \textbf{\bibinfo{volume}{A720}}, \bibinfo{pages}{399} (\bibinfo{year}{2003}),
  \eprint{hep-ph/0206219}.

\bibitem[{\citenamefont{Epelbaum et~al.}(2000)\citenamefont{Epelbaum,
  Gl{\"o}ckle, and Mei{\ss}ner}}]{Epelbaum:1999dj}
\bibinfo{author}{\bibfnamefont{E.}~\bibnamefont{Epelbaum}},
  \bibinfo{author}{\bibfnamefont{W.}~\bibnamefont{Gl{\"o}ckle}},
  \bibnamefont{and} \bibinfo{author}{\bibfnamefont{U.-G.}
  \bibnamefont{Mei{\ss}ner}}, \bibinfo{journal}{Nucl. Phys.}
  \textbf{\bibinfo{volume}{A671}}, \bibinfo{pages}{295} (\bibinfo{year}{2000}),
  \eprint{nucl-th/9910064}.

\bibitem[{\citenamefont{Zemp}(2004)}]{Zemp}
\bibinfo{author}{\bibfnamefont{P.}~\bibnamefont{Zemp}}, Ph.D. thesis,
  \bibinfo{school}{University of Berne} (\bibinfo{year}{2004}).

\bibitem[{\citenamefont{Beane and Savage}(2003)}]{Beane:2002aw}
\bibinfo{author}{\bibfnamefont{S.~R.} \bibnamefont{Beane}} \bibnamefont{and}
  \bibinfo{author}{\bibfnamefont{M.~J.} \bibnamefont{Savage}},
  \bibinfo{journal}{Nucl. Phys.} \textbf{\bibinfo{volume}{A717}},
  \bibinfo{pages}{104} (\bibinfo{year}{2003}), \eprint{nucl-th/0204046}.

\bibitem[{\citenamefont{Borasoy and Grie{\ss}hammer}(2003)}]{Borasoy:2003gf}
\bibinfo{author}{\bibfnamefont{B.}~\bibnamefont{Borasoy}} \bibnamefont{and}
  \bibinfo{author}{\bibfnamefont{H.~W.} \bibnamefont{Grie{\ss}hammer}},
  \bibinfo{journal}{Int. J. Mod. Phys.} \textbf{\bibinfo{volume}{E12}},
  \bibinfo{pages}{65} (\bibinfo{year}{2003}).

\bibitem[{\citenamefont{Mei{\ss}ner et~al.}(2005)\citenamefont{Mei{\ss}ner,
  Raha, and Rusetsky}}]{Meissner:2005bz}
\bibinfo{author}{\bibfnamefont{U.-G.} \bibnamefont{Mei{\ss}ner}},
  \bibinfo{author}{\bibfnamefont{U.}~\bibnamefont{Raha}}, \bibnamefont{and}
  \bibinfo{author}{\bibfnamefont{A.}~\bibnamefont{Rusetsky}},
  \bibinfo{journal}{Eur. Phys. J.} \textbf{\bibinfo{volume}{C41}},
  \bibinfo{pages}{213} (\bibinfo{year}{2005}), \eprint{nucl-th/0501073; Erratum
  (Eur. Phys. J. C, to be published)}.

\bibitem[{\citenamefont{Nogga and Hanhart}(2005)}]{Nogga:2005fv}
\bibinfo{author}{\bibfnamefont{A.}~\bibnamefont{Nogga}} \bibnamefont{and}
  \bibinfo{author}{\bibfnamefont{C.}~\bibnamefont{Hanhart}}
  (\bibinfo{year}{2005}), \eprint{nucl-th/0511011}.

\bibitem[{\citenamefont{Gasser et~al.}(2003)\citenamefont{Gasser, Rusetsky, and
  Scimemi}}]{Gasser:2003hk}
\bibinfo{author}{\bibfnamefont{J.}~\bibnamefont{Gasser}},
  \bibinfo{author}{\bibfnamefont{A.}~\bibnamefont{Rusetsky}}, \bibnamefont{and}
  \bibinfo{author}{\bibfnamefont{I.}~\bibnamefont{Scimemi}},
  \bibinfo{journal}{Eur. Phys. J.} \textbf{\bibinfo{volume}{C32}},
  \bibinfo{pages}{97} (\bibinfo{year}{2003}), \eprint{hep-ph/0305260}.

\bibitem[{\citenamefont{Mei{\ss}ner}(2005)}]{Meissner:2005ba}
\bibinfo{author}{\bibfnamefont{U.-G.} \bibnamefont{Mei{\ss}ner}},
  \bibinfo{journal}{Proc. Sci.} \textbf{\bibinfo{volume}{LATT2005}},
  \bibinfo{pages}{009} (\bibinfo{year}{2005}), \eprint{hep-lat/0509029}.

\bibitem[{\citenamefont{Lyubovitskij et~al.}(2002)\citenamefont{Lyubovitskij,
  Gutsche, Faessler, and Vinh~Mau}}]{Lyubovitskij:2001zn}
\bibinfo{author}{\bibfnamefont{V.~E.} \bibnamefont{Lyubovitskij}},
  \bibinfo{author}{\bibfnamefont{T.}~\bibnamefont{Gutsche}},
  \bibinfo{author}{\bibfnamefont{A.}~\bibnamefont{Faessler}}, \bibnamefont{and}
  \bibinfo{author}{\bibfnamefont{R.}~\bibnamefont{Vinh~Mau}},
  \bibinfo{journal}{Phys. Rev.} \textbf{\bibinfo{volume}{C65}},
  \bibinfo{pages}{025202} (\bibinfo{year}{2002}), \eprint{hep-ph/0109213}.

\end{thebibliography}

\end{document}